\documentclass[twocolumn,english,aps,prx,reprint,superscriptaddress,showpacs,longbibliography]{revtex4-1}
\usepackage{amsmath,amsfonts,amsthm,amssymb,mathrsfs,bbm}
\usepackage{graphicx,mathptmx}
\usepackage[colorlinks=true,linkcolor=blue,citecolor=blue,breaklinks=true]{hyperref}
\usepackage{subfigure}
\usepackage{epstopdf}
\usepackage{color}
\usepackage{array}
\usepackage{multirow,booktabs,dcolumn}
\usepackage{filecontents}
\usepackage{algorithm}
\usepackage{algpseudocode}
\usepackage{enumitem}   
\usepackage{soul}
\usepackage{ulem}

\makeatletter
\def\arrvline{\hfil\kern\arraycolsep\vline\kern-\arraycolsep\hfilneg}
\makeatother

\begin{document}
%\begin{document}
\title{ Precursors-driven machine learning prediction of chaotic extreme pulses in Kerr resonators}

\author{S. Coulibaly}
\email{saliya.coulibaly@univ-lille.fr}	
\affiliation{Universit\'e de Lille, CNRS, UMR 8523 - PhLAM - Physique des Lasers Atomes et Mol\'ecules, F-59000 Lille, France.}
\author{F. Bessin}
\affiliation{Aston Institute of Photonics Technologies, Aston University, Birmingham B4 7ET, UK.}
\author{M. G. Clerc}
\affiliation{Departamento de F\'{i}sica {and Millennium Institute for Research in Optics}, FCFM, Universidad de Chile, Casilla 487-3, Santiago, Chile.}
\author{A. Mussot}
\affiliation{Universit\'e de Lille, CNRS, UMR 8523 - PhLAM - Physique des Lasers Atomes et Mol\'ecules, F-59000 Lille, France.}

%\newpage
\begin{abstract}
%\begin{abstract}%144 words --- 1039 characts (requested: 150--If Letter provide references) \;

%-------------Statistics-----------------%
% Abstract: 139  [150 Requested]
% Main Text: 1364  [Requested Letter 1500: Article 3000]
%Method: 
% References: 22  [Requested: Letter 30 - Article 50]
% Figures: 4  [Requested: Letter 5 - Article 6]
%-------------Statistics-----------------%

{Machine learning algorithms have opened a breach in the fortress of the prediction of high-dimensional chaotic systems. 
Their ability to find hidden correlations in data can be exploited to perform model-free forecasting of spatiotemporal chaos 
and extreme events. However, the extensive feature of these evolutions constitutes a critical limitation for full-size forecasting processes. Hence, the main challenge for forecasting relevant events is to establish the set of pertinent information. Here, we identify precursors 
 from the transfer entropy of the system and a deep Long Short-Term Memory network to forecast the complex dynamics of a system evolving in a high-dimensional spatiotemporal chaotic regime. Performances of this triggerable model-free prediction protocol based on the information flowing map are tested from  experimental data originating from a passive resonator operating in such a complex nonlinear regime. 
We have been able to predict the occurrence of extreme events up to 9 round trips after the detection of precursor, {\textit{i.e.}, }
3 times the horizon provided by Lyapunov exponents, with 92 $\%$ of true positive predictions leading to 60 $\%$ of accuracy.}
\end{abstract}

\date{\today}
%Resonator - fiber optics, 42.81.Qb
%Nonlinear optics, 42.65.-k
%Kerr effect nonlinear optics, 42.65.Hw
%Pattern formation in complex systems, 89.75.Kd
%General theory of critical region behavior, 64.60.fd
%Multicritical points, 64.60.Kw
%Numerical simulations Chaos, 05.45.Pq
%Optical Chaos, 42.65.Sf
%High-dimensional Chaos, 05.45.Jn
\maketitle

%\end{abstract}
\section{Introduction}

Large-aspect-ratio deterministic systems operating out of equilibrium can become extremely sensitive to {the} initial conditions 
when undergoing chaotic spatiotemporal evolution \cite{Poincare1908,Lorenz1963,Ruelle1982,Yorke1975}. Spatiotemporal 
chaos may be understood as the exponential destruction of information in both time and space, making the dynamics 
require many spatially distributed chaotic elements to be described \cite{Cross1994}. With these elements, accurate 
{modeling} of such a system lies in two key points---a good description of the physical equations and minimal 
uncertainty in the initial conditions.  Despite many years of intensive research to understand the 
complex dynamics of chaos,  most are limited to theoretical investigations. Only a few experimental   works had been reported due to the huge precision required to gain knowledge on the initial conditions.
Recently, improvements of supervised machine learning algorithms have brought new perspectives for the forecasting of spatiotemporal complex dynamics {in optics  \cite{genty_machine_2021}, economy \cite{ghoddusi2019machine}, power grid load \cite{rudin2011machine}, and meteorology \cite{vlachas2018data,ham2019deep,li2020deep}, to mention a few. These studies were performed mainly using deep  learning, recurrent, and echo state networks}. By providing model free 
processes it could be possible that chaos theory tools are no more necessary to handle time series in general. 
Even though powerful, machine learning based forecasting can present major interest when dealing with 
spatiotemporal chaos. Indeed, the specificity of this chaos is its extensive  feature; 
that is, more the larger the system, more the larger the total number of  {the coupled} nodes in the network.  This makes the problem rapidly unsolvable for high-dimensional spatiotemporal chaotic systems.  Thus, alternative strategies based on local intensive order parameters other than predicting the whole system are needed.

Here, we propose a demonstration that model-based and model-free tools can be combined to provide triggerable 
local forecasting of the extreme events in chaotic regimes.  
Namely, the forecasting process is activated when relevant information is identified. 
Answering the question of \textit{when} and \textit{where} the extreme events will emerge, 
we also address the question of \textit{what is coming?} that is, what will be the profile of the coming event. 

\section{The experimental setup}

\begin{figure}[H]
	\centering
	\includegraphics[width=0.5\textwidth]{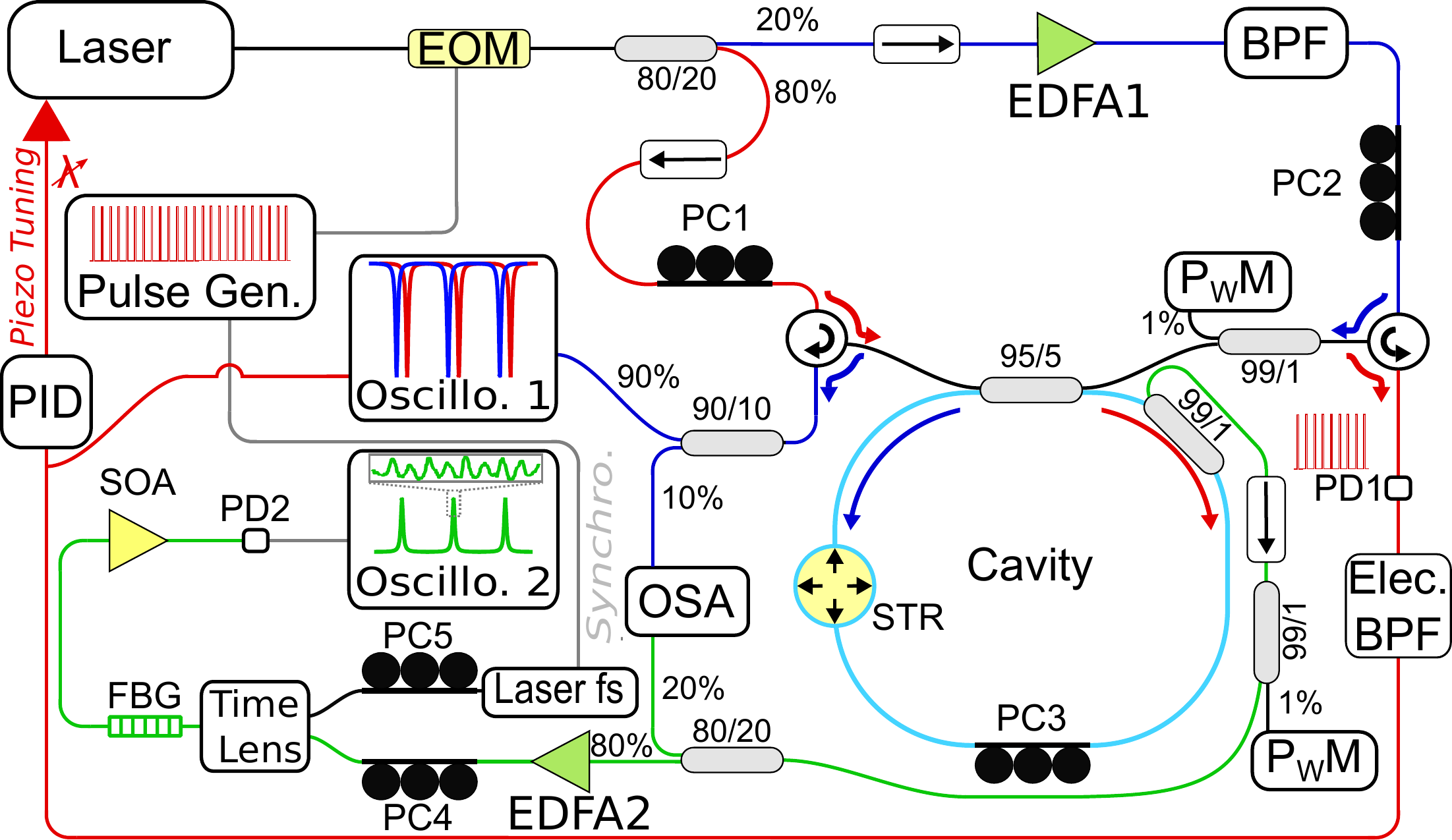}
	\caption{Experimental setup. P$_W$M, powermeter; PC$_{1-5}$, polarization controller; PD$_{1-2}$, photodetector; OSA, optical spectrum analyzer; Elec. BPF: electronic band-pass filter; Oscillo.$_{1-2}$, oscilloscope; Pulse Gen., pulse generator; EOM, electro-optic modulator; PID, proportional-integrate-derivative; EDFA$_{1-2}$, erbium-doped fiber amplifier; BPF, band-pass filter; STR, fiber stretcher; FBG, fiber Bragg grating; SOA, semiconductor optical amplifier; Laser fs, femtosecond laser.}
	\label{fig:setup}
\end{figure}

A details sketch of the experimental setup is depicted in Fig. \ref{fig:setup}. It is similar to those used in Ref. \cite{bessinP1P2} . It consists of a passive fiber ring cavity build with a specially designed dispersion shifted fiber ($\beta_{DSF}=-3.8$  ps$^2$/km at $1545$ nm and $\gamma_{DSF}=2.5$ W$^{-1}$.km$^{-1}$) closed by a 95/5 coupler made of the same fiber to get a perfectly uniform cavity of $132.9$ meter-long with a finesse of $15.6$. We drive the cavity with a train of square shaped pulses of $1$ ns duration. This configuration prevents from stimulated Brillouin scattering and to generate high pick power to trigger the parametric process. These pulses are generated from a continuous wave (cw) laser at $1545$ nm (with a narrow linewidth, less than $100$ Hz) whose intensity is chopped by an electro-optic modulator (EOM). The repetition rate of these pulses is set to match with the repetition rate of the cavity, in order to drive the system synchronously and get one pulse per roundtrip. Pulses are then amplified by an erbium doped amplifier and filtered out by a thin bandwidth filter (BPF, $100$ GHz) to remove amplified spontaneous emission (ASE) in excess. Finally, pump pulses are launched into the cavity through the right port of the cavity propagating in the anticlockwise direction (blue arrows). Note that, we added a 99/1 tap coupler just before the input port of the cavity for input power monitoring and setting. Due to the interferometric nature of such a system the linear phase accumulated by pump is extremely sensitive to external perturbations (change in pressure, temperature) and need to be stabilized. For this purpose, a fraction of the output power of the EOM is launched through the left port of the cavity, propagating in the clockwise direction (red arrows). This weak signal detected at the cavity output by a photodetector (PD1) provides an error signal for a feedback loop system (proportional-integrate-derivative) which finely tunes the cw laser wavelength. As in Ref. \cite{coen1997modulational, bessinP1P2}, a combination of three polarization controllers and measurements of a fraction of cavity output signals are used to control the cavity detuning (normalized detuning set to $\Delta=1.1$, monostable \cite{coen1999bistable}). In order to study the intracavity field, we added just before the coupler closing the cavity a 99/1 tap coupler. A part of this extracted field (20\%) is analyzed by means of optical spectrum analyzer while the other part (80\%) is amplified by a low noise EDFA and then studied by a commercial time-lens (Picoluz ultra-fast temporal magnifier, Thorlabs) based on the results published in Ref. \cite{salem2013application}. 

The time stretching effect was obtained by pumping the time-lens with a femtosecond laser centered at $1570$ nm providing pulses with a fixed repetition rate of $99.88$ MHz. This laser was used as a reference clock for the EOM such as the repetition rate of cavity pump pulses is an exact multiple of the femtosecond laser (typically $65$ times in our case). In order to drive the cavity in a perfectly coherent way, we added a stretcher inside the system, thus we could finely tuned the cavity length such as the pump pulses repetition rate matched with the cavity repetition rate. The magnified signal (magnified factor of $57$) was filtered by means of a fiber Bragg grating to perfectly remove the femtosecond pump in excess, and then slightly amplified by a semiconductor optical amplifier to be recorded by a fast photodiode and an oscilloscope (70 GHz bandwidth each). Thanks to this time-lens, we were able to record at each round-trip the intracavity temporal pattern over a window of $36$ ps with a resolution of about $300$ fs (shorter than the local dynamics time scale). {We will use these data either to train the network or to test its performances.}

\section{Complex dynamics characterization and information transfer}
\begin{figure}[h]
\begin{centering}
\includegraphics[width=.5\textwidth]{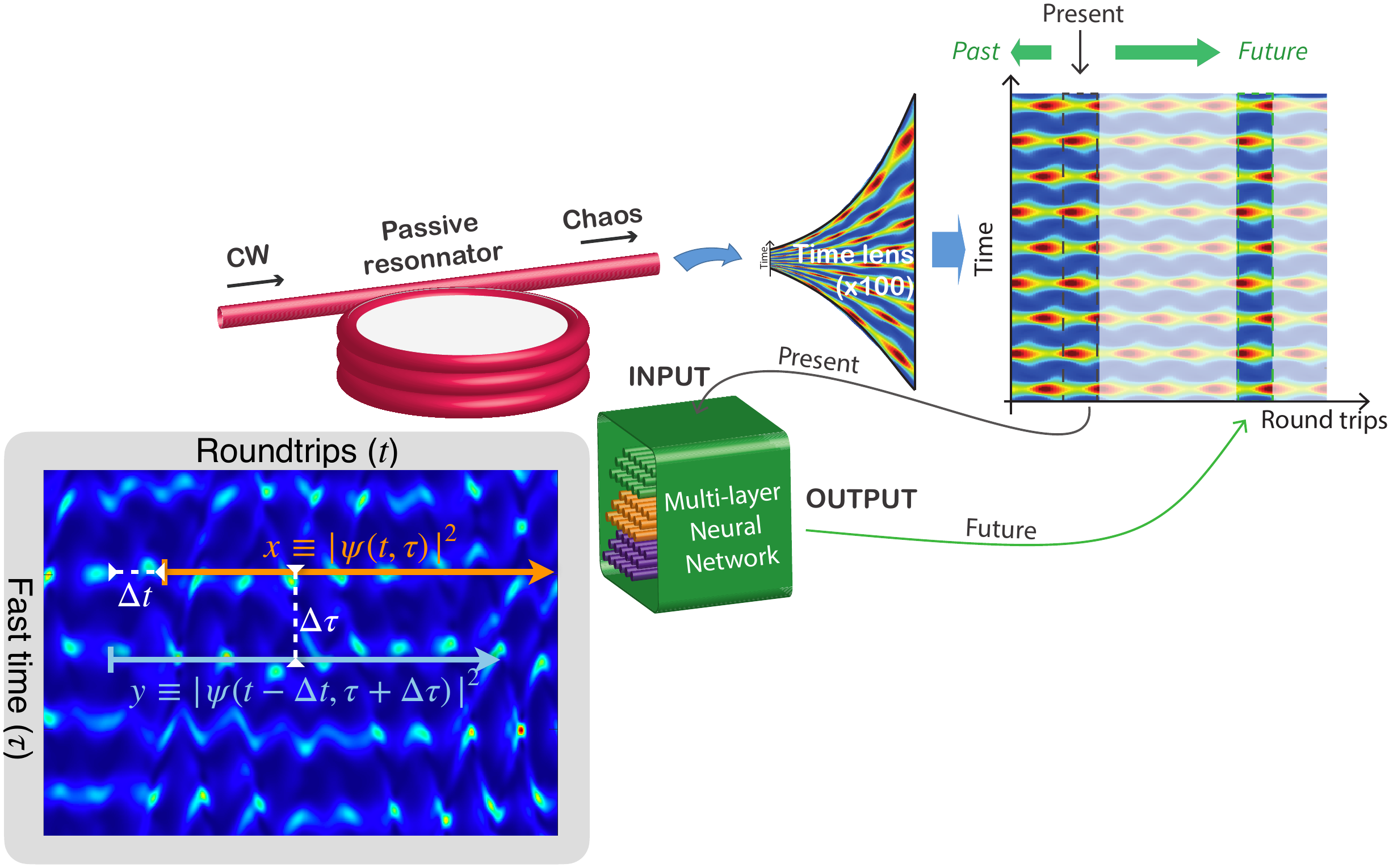}
\par\end{centering}
\caption{\label{fig1:cartoon}  Schematic representation of prediction method of the spatiotemporal chaos at the output of the 
optical fiber ring synchronously pump close to a resonance frequency. CW: continuous wave. { In the grey panel, we illustrate how the signals are selected to compute the transfer entropy map. The details of the multi-layer Network are given in the Appendix~\ref{ap:predprotocol} section.}}
\end{figure}

The degree of sensitivity to the initial {conditions} is formally deduced from the value of the largest Lyapunov exponent (LE). 
 This exponent can be computed for low dimensional systems based on the mathematical model 
if known \cite{ott2002chaos} or  from the time series \cite{wolf1985determining}. However, for extended systems, 
the main characteristics of the chaos require the knowledge of the continuous spectrum of LEs~\cite{skokos2010lyapunov}. 
Hence, from experimental data, only the consequences of the chaos can be measured and not their  analytical characteristics. 
Indeed, LE is also interpreted as the production
rate of entropy during the evolution. Likewise,
in high-dimensional chaos, correlation ranges are much smaller than the actual size of the system.
Consequently, according to information theory \cite{cai2001spatiotemporal}, mutual information between two {locations} $x_1$ 
and $x_2$ may exponentially vanish  as the separation $||x_1-x_2||\rightarrow \infty$. For a system composed of two 
signals $x$ and $y$ with joint probability $p(x,y)$ the Shanon entropy~\cite{shannon1949mathematical} is 
$H_1=-\sum_{x,y}p(x,y)\log[p(x,y)]$. If the same processes are supposed independent it comes $H_2=-\sum_{x,y}p(x,y)\log[p(x)p(y)]$. 
The mutual information is then $I_{XY}=H_2-H_1=\sum_{x,y}p(x,y)\log\left[\frac{p(x,y)}{p(x)p(y)}\right]$. { 
To determine  which of the two signals provide more  information regarding its own past, it is useful to compute 
the transfer entropy (TE) \cite{schreiber2000measuring,lizier2008local,wibral2013measuring,abdul2014quantifying,boba2015efficient} : $TE_{Y\rightarrow X}=\sum_{x,y}p(x_{n+1},x_n^h,y^h_n)\log\left[\frac{p(x_{n+1}|x_n^h,y_n^h)}{p(x_{n+1}|x_n^h)}\right]$, with $n$ the current iteration and $h$ the history length. Hence, taking $x$ and $y$ 
as the measured data at different locations separated by $\Delta t$ (slow time in Fig.~\ref{fig1:cartoon}) 
and lagging one over the other by $\Delta \tau$ (fast time in Fig.~\ref{fig1:cartoon}), one can construct the map $TE_{Y\rightarrow X}\left(\Delta t,\Delta \tau\right)$ or $TE_{X\rightarrow Y}\left(\Delta t,\Delta \tau\right)$ as sketched in Fig.~\ref{fig1:cartoon}. With the two-point correlation length (see  Supplementary  Materials), 
TE will be the model-free tool that we will use to 
measure the impact of the spatiotemporal chaos in our {system}. } 
{ In practice, there are a many codes that allow to compute the transfer entropy of continuous time series. Here, for our transfer entropy maps, we have used the open source  JIDT software package\cite{lizier2014jidt} (https://github.com/jlizier/jidt/). The portability of this JIDT Java-based code, with no installation requirement have motivated our choice.
}
\section{Spatiotemporal chaos in an optical fiber ring resonator}
{ Figure~\ref{fig1:cartoon} sketches up the prediction protocol of chaotic extreme pulses in a Kerr resonator}. 
The data are obtained from a passive resonator made of an optical fiber ring synchronously  pumped close to a cavity resonance (see Methods and supplemental document for details). The repetition rate of the cavity is about $1.54$ MHz ($64.9$ $\mu s$) and the local dynamics time scale { is of the} order of the ps. { For simplicity}, the ring was set to operate in a monostable regime, \textit{i.e.}, a region where the transmission function is single valued for a given pump power. {By pumping the cavity well above the cavity threshold, typically a few times, } the continuous wave solution breaks into a periodic wave train, which in turn experiences an oscillatory instability and then evolves onto a chaotic regime \cite{coulibaly2019turbulence, pasquazi_micro-combs_2018}. {This} current sequence is {universal and can be observed in many other fields of physics \cite{Cross1994, ott2002chaos}}. 
\begin{figure}[h]
\begin{centering}
\includegraphics[width=.4\textwidth]{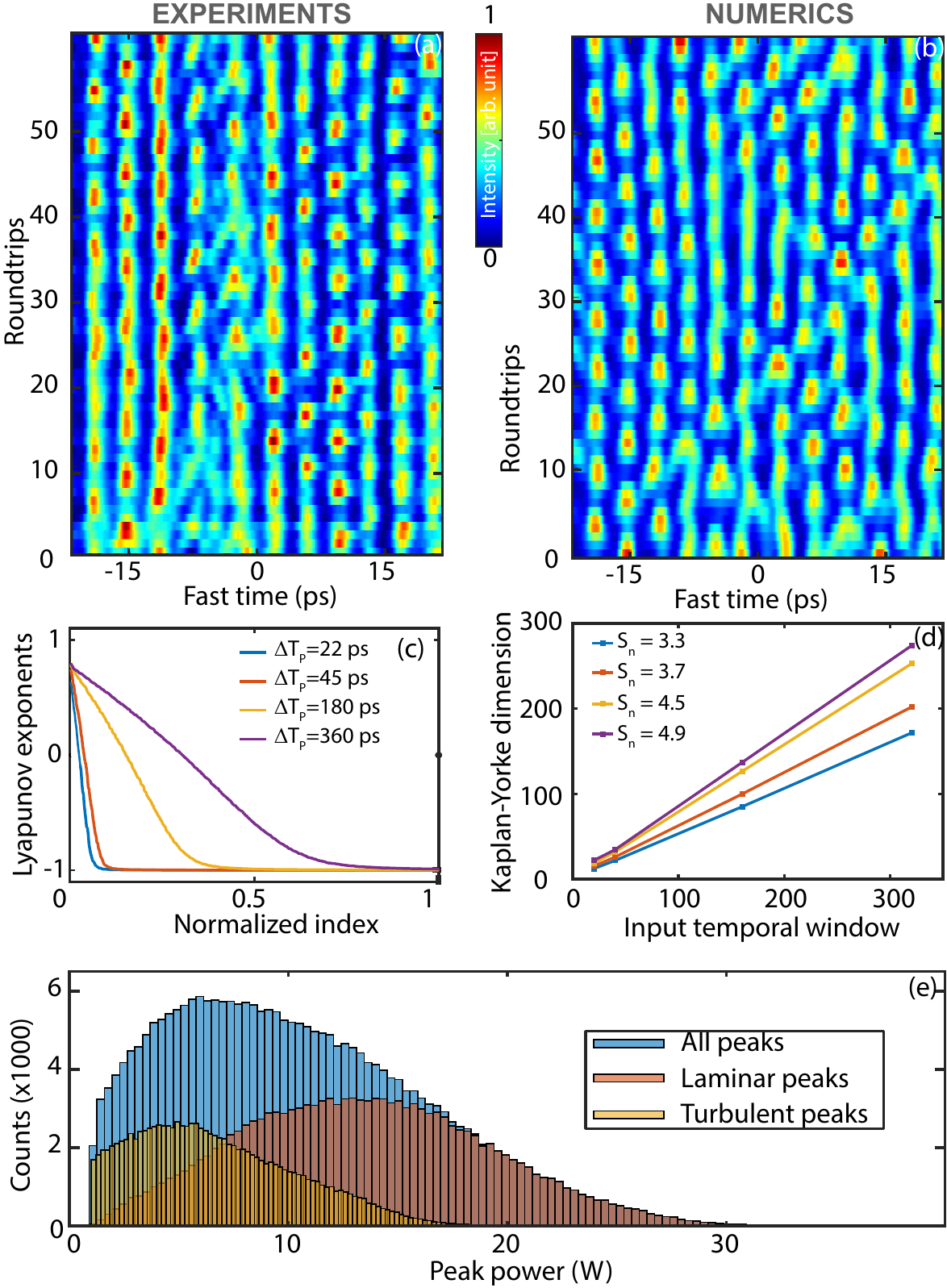}
\par\end{centering}
\caption{\label{fig2} Typical spatiotemporal turbulent dynamics (a) from experiments and (b) from numerics 
(LLE, Eq.~{(S7)} in supplemental documents). (c) Lyapunov spectrum for different time windows ($\Delta T_P$). 
(d) Kaplan-Yorke dimension as function of the temporal window ($\Delta T_P$) for different output power normalized to 
the cavity threshold ($S_n$). All parameters are listed in Appendix~\ref{ap:numerics}.  {(e) Probability density functions 
of the all the peaks (blue), laminar peaks (orange) and turbulent peaks (yellow) from numerics for a detection threshold 
set at the mean value of the intracavity power. The evolution of these distributions is provided as an additional material (moviepdf.gif).}}
\end{figure}
{The dynamics of} the light circulating in the cavity is accurately modelled by the driven and damped nonlinear Schr\"odinger equation \cite{haelterman1992dissipative} (see Appendix~\ref{ap:numerics}), referred to as 
{ the} Lugiato-Lefever equation {(LLE)} \cite{lugiato1987spatial}. The LLE has the advantage that we can  {use both model-based and model-free tools to} compute all the quantities needed to {characterize} the spatiotemporal complexity. 
\begin{figure*}
\begin{centering}
\includegraphics[width=1\textwidth]{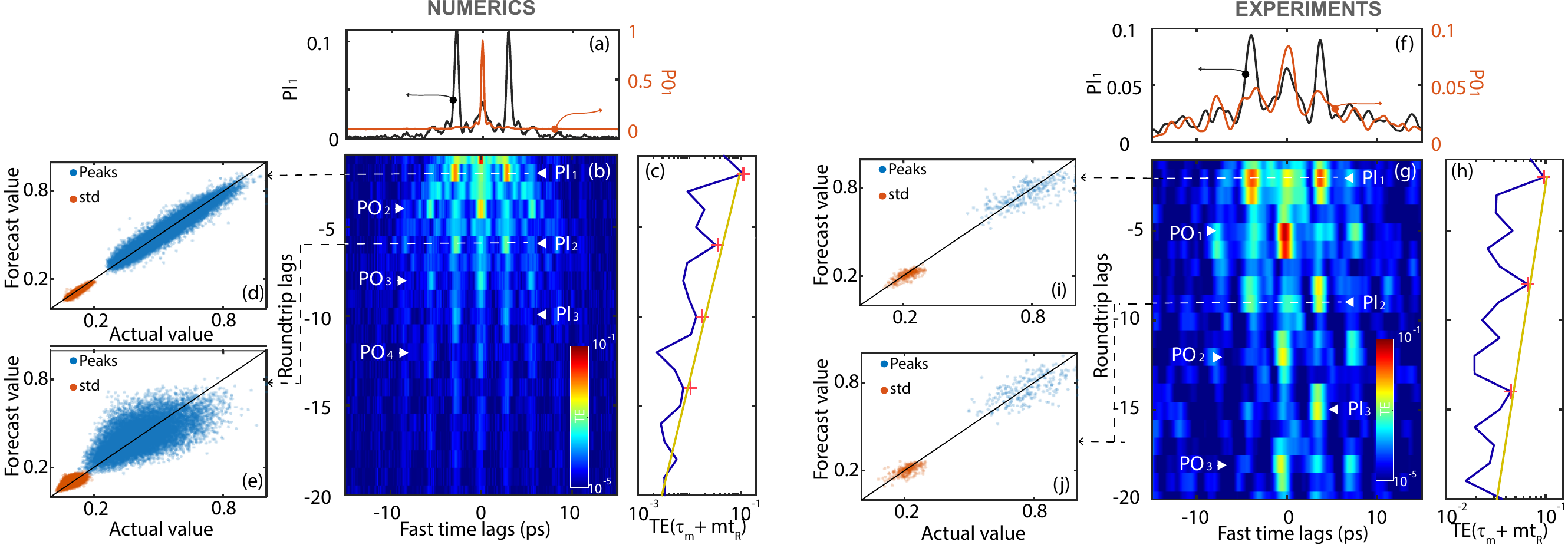}
\par\end{centering}
\caption{\label{fig4:TE} 
{ (a)-(e) Numerical simulations of the LLE (Eq.~(\ref{eq:LL}) in Appendix~\ref{ap:numerics}) and (f)-(j) from experiments in a Kerr resonator. (b) and (g) show the 2D plot of the transfer entropy. (a) and (f) represent profiles of P0$_1$ and P1$_1$. The evolution of the transfer entropy at the fast time lag given by the P1$_1$ maximum with respect to the roundtrip lags are shown in (c) and (h). The blue solid line shows the evolution of the transfer entropy as a function of the roundtrips, the symbols (+) mark peaks at each P1$_n$,  and the strait yellow line the best exponential fit from these maxima. (d) and (e) show correlation maps after the supervised training using the association between detected pulses and their P1$_1$ and P1$_2$ precursors respectively from numerical simulations and (i) and (j) from experiments.  In panels (d), (e), (i), and (j) the horizontal axes, \textit{Actual value},  stands for the measured peak value (blue points) and the standard deviation of the observed pulses (red points). The vertical axes, \textit{Forecast value}, accounts for the predicted peak value and their standard deviation.}}
\end{figure*}
Figure~\ref{fig2}(a) shows an example of the complex { behavior} 
obtained experimentally by pumping the cavity well above the nonlinear threshold (3 times the emission threshold). 
{It illustrates the output cavity field in the time domain, round trip to round trips. An almost periodic pulse train about 3.8 ps 
period with pulse duration of typically 1.8 ps can be observed. Pulse positions and shapes modifications in this 
two dimensional map is characteristic of a spatiotemporal chaos \cite{coulibaly2019turbulence}. 
We performed numerical simulations with experimental parameters. They are depicted in Fig.~\ref{fig2}(b) 
and look similar to experimental results in Fig.~\ref{fig2}(a).} { The fine characterization of the complexity of 
this spatiotemporal chaotic regime had been performed from standard analysis tools \cite{ott2002chaos} 
either by changing the time window or the pump power. {Firstly, Fig.~\ref{fig2}(c) shows the Lyapunov 
spectrum evolution for different time window durations ($\Delta T_p$) for a pump power set to $S_n=4.9$ 
(about 5 times the nonlinear threshold). The spectrum broadens by increasing the time window, which is a clear 
signature of a spatiotemporal chaos. Secondly, Fig.~\ref{fig2}(d) represents Kaplan-Yorke dimension evolution 
as a function of the temporal window for several pump powers ranging from 3.3 to 4.9 times the cavity threshold. 
The slope of {curves} increases with the pump power that confirms the spatiotemporal chaotic nature of the process. 
More precisely, theses slopes provide} an estimation of the duration $\Delta T_{stc}$ of independent chaotic subdomains. 
It is of  the order of 1 ps in this 
case and much smaller ($\Delta T_{stc}\ll\Delta T_p$) than the time widow duration (36 ps here, see Figs.~\ref{fig2}(a) and (b)). 
Lyapunov spectra also {enable to estimate} the production rate of information during {evolution along the slow 
time (cf. Fig.\ref{fig1:cartoon}}). For high-dimensional chaos the mean metric entropy corresponds to the 
Kolmogorov-Sinai entropy $h_{KS}=\sum_{\lambda_i>0}\lambda_i$ \cite{wei2000quantifying,shaw1981strange}. 
The fluctuations lifetime over the cavity roundtrips is given by $\tau_{stc}=1/h_{KS}$. From experimental 
parameters (see Table~S1 in the Supplementary material) we found that $\tau_{stc}<t_R$ where $t_R$  represents the cavity roundtrip,  which is a key feature of a system evolving 
{into} a high dimensional chaotic regime.  { On the other hand,  the description of the spatiotemporal chaos can be achieved} { by analogy with hydrodynamics \cite{coulibaly2019turbulence}. The dynamics is an irregular succession of laminar and turbulent flows. A detailed statistical study of the laminar or turbulent domains was 
 performed in \cite{coulibaly2019turbulence},  in which the probability distribution of the laminar/turbulent domains has  the following mixture function: $P(x)=(Ax^{-\mu}+B)e^{-mx}$. All the constants depend only on the parameters except $m$ which also changes with the value of the power set to separate the laminar and turbulent domains. The burst detected during the evolution can be labelled according to their location in a laminar or a turbulent flow, respectively. Distributions of all the bursts, 
 those located in laminar, and turbulent domains are shown in {Figs.~\ref{fig2}(e) for a threshold set at the mean value the intracavity power. Highest bursts are mainly located in the laminar flows, and it is even more pronounced for highest threshold values (see movie in supplemental material for other thresholds)}. Hence, we can use the transfer entropy to map the information flow from the neighborhood and the past. To this end, we compute $TE_{Y\rightarrow X}\left(\Delta t,\Delta \tau\right)$ 
 with $X\equiv\left|\psi(t,\tau)\right|^2$ and $Y\equiv\left|\psi(t-\Delta t,\tau+\Delta\tau)\right|^2$, $\psi$ being the considered field.} {
 \section{Estimation of the local dynamics range and predictibility}
The qualitative feature of a high dimensional complex behavior is the finite nature of the interaction range. Theory of dynamical systems and information theoretic have provided various estimations of the this range. In a spatiotemporal chaotic regime, the equal time two-point correlation range and the Lyapunov dimension can be useful estimators. 
\subsection{Spatiotemporal chaos dimension $\xi_\delta$}\ \\
The Kaplan-Yorke dimension grows linearly with the volume of a high dimensional chaotic system. For all parameters fixed, is worth to provide an intensive  characterization of the chaoticity level. This can be done by computing the slope of the Kaplan-Yorke dimension curve with respect to the volume \cite{coulibaly2019turbulence}. The inverse of this slope  estimates the size of the independent subdomains produced by the presence of the attractor. Figure~\ref{fig:loc_range}(a) shows the typical evolution of $\xi_\delta$ with respect of the pump power in the LLE. It can be seen that the range of independent subdomains decrease with pumping. In the fully developed turbulent regime ($S_n/S_{tn}>9$) parameters we have $\xi_\delta\ll\Delta T_p$.
 \begin{figure*}
\centering
\includegraphics[width=0.9\textwidth]{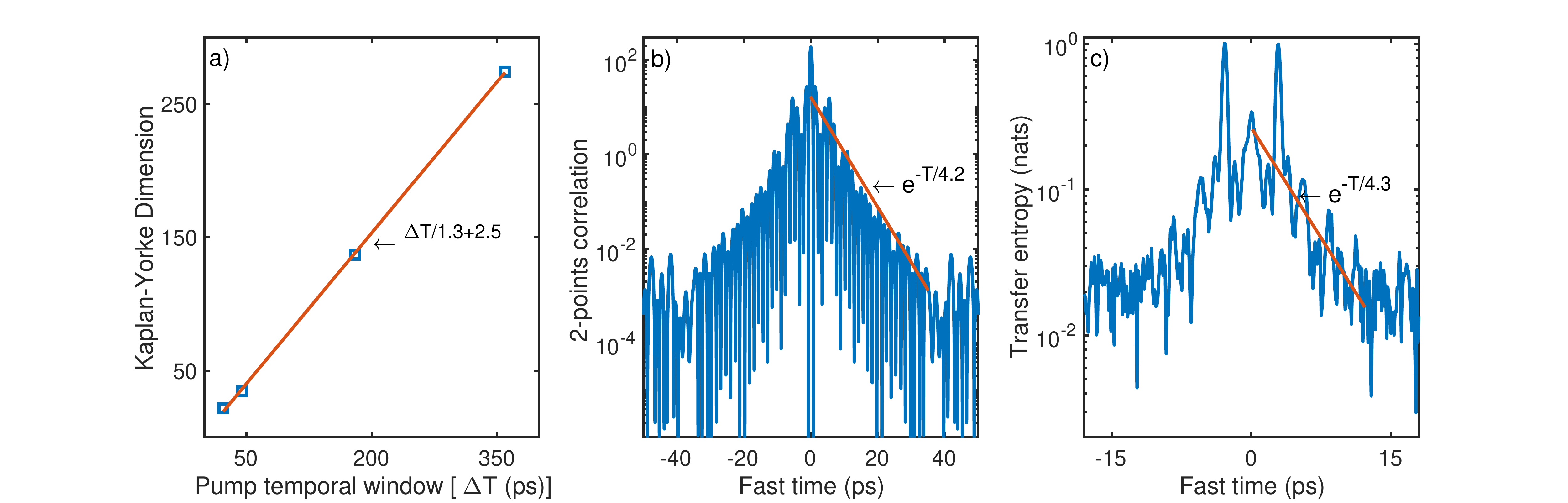}
	\caption{Local dynamics range of the LLE equation for $\Delta=1.1$ and $S=4.9$. (a) Lyapunov dimension density, (b) Correspond to the equal time two-points correlation and, (c) the transfer entropy vanishing range. Red lines in (b) and (c) are the linear trend extracted from the data.}
	\label{fig:loc_range}
\end{figure*}
\subsection{Equal time correlation dimension $\xi_2$}\ \\
In the complex evolution the probability of two locations  separated by $\delta T$ to behave coherently is given obtained by computing the function: \cite{Ohern1996lyapunov,egolf1994relation,cross1993pattern}:
\begin{equation}
\label{eq:2ptcorr}
C\left(\delta T\right)=\langle\left(\psi\left(\delta T+T^\prime,t\right)-\langle\psi\rangle\right)\left(\psi\left(T^\prime,t\right)-\langle\psi\rangle\right)\rangle.
\end{equation}
The brackets  $\langle\cdot\rangle$ stand for the average process. $C\left(\delta T\right)$ is the equal time two point correlation function. The computation cost of this function is generally reduced by using Wiener-Khintchin theorem \cite{reif1965fundamentals,egolf1995characterization}. The correlation length $\xi_2$ is defined as the exponential decay of $C\left(\delta T\right)$. For the set of parameters used here, the correlation function shown in Fig.~\ref{fig:loc_range}(b). We found that the $\xi_2\simeq4.2$ ps, which is much larger than $\xi_\delta=1.3$ ps.

\begin{figure}
\centering
\includegraphics[width=.45\textwidth]{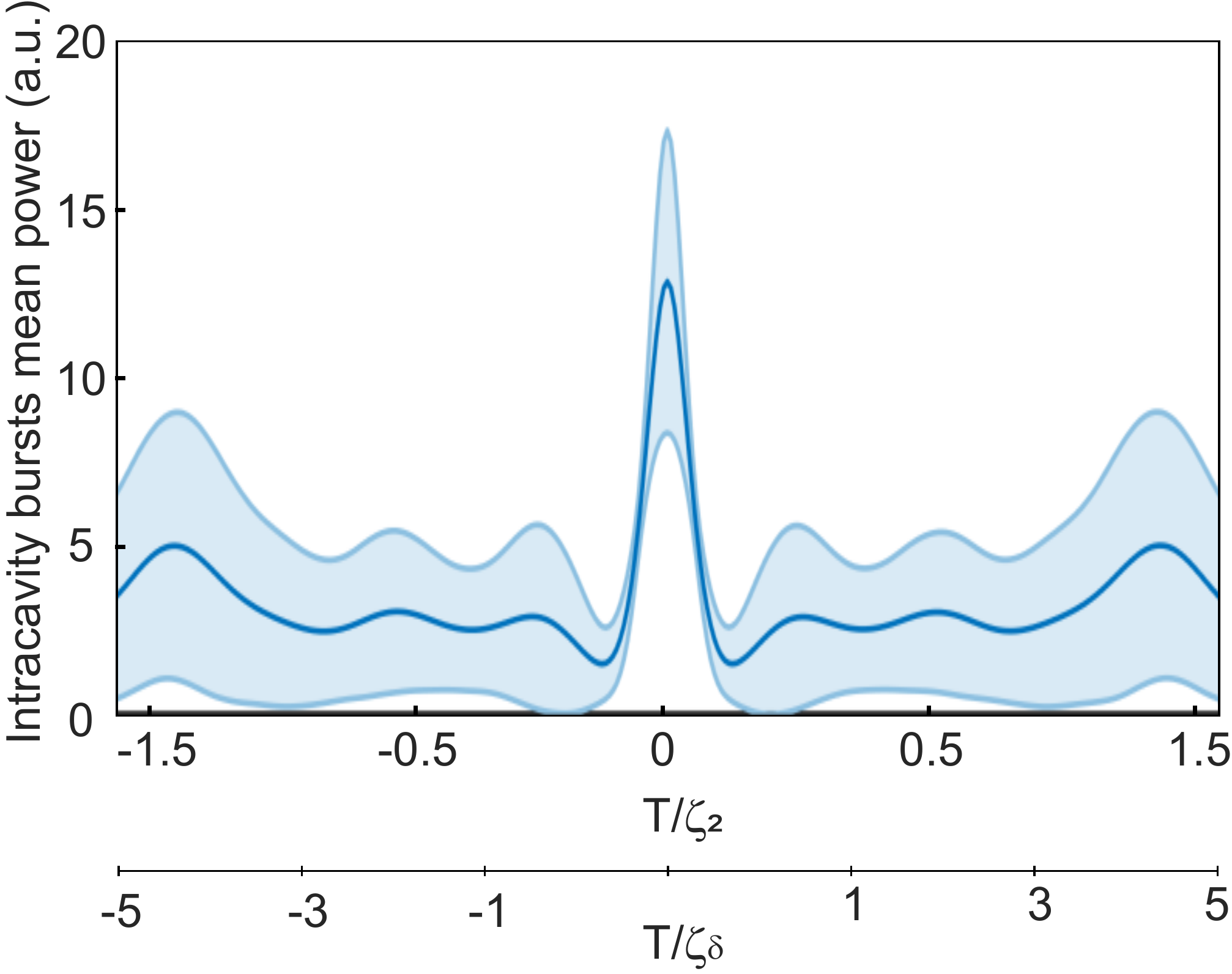}
	\caption{Mean profile of the intensity bursts with the standard deviation range (shaded region) in the unit of  $\xi_\delta$ and $\xi_2=\xi_I$ for $\Delta=1.1$ and $S=4.9$.}
	\label{fig:range_comp}
\end{figure}

The direct determination of $C\left(\Delta\tau\right)$ is quite costly in calculation time. However, by using the Wiener-Khintchin theorem \cite{reif1965fundamentals,egolf1995characterization}, it is computed by the following process: first time-averaging the Fourier spectra and next taking the inverse Fourier transform {of} its magnitude squared. Since the experimental spectra result from an averaging process over a large number of cavity roundtrip, $C\left(\Delta\tau\right)$ can also be computed taking the inverse Fourier transform of the measured spectrum. 
Hence, for the LL equation (1) %{(\ref{eq:LL})}
, we have computed $\xi_\delta$ end $\xi_2$ with respect to the input pump intensity. 
%\subsection{}
\begin{figure}
\centering
\includegraphics[width=.45\textwidth]{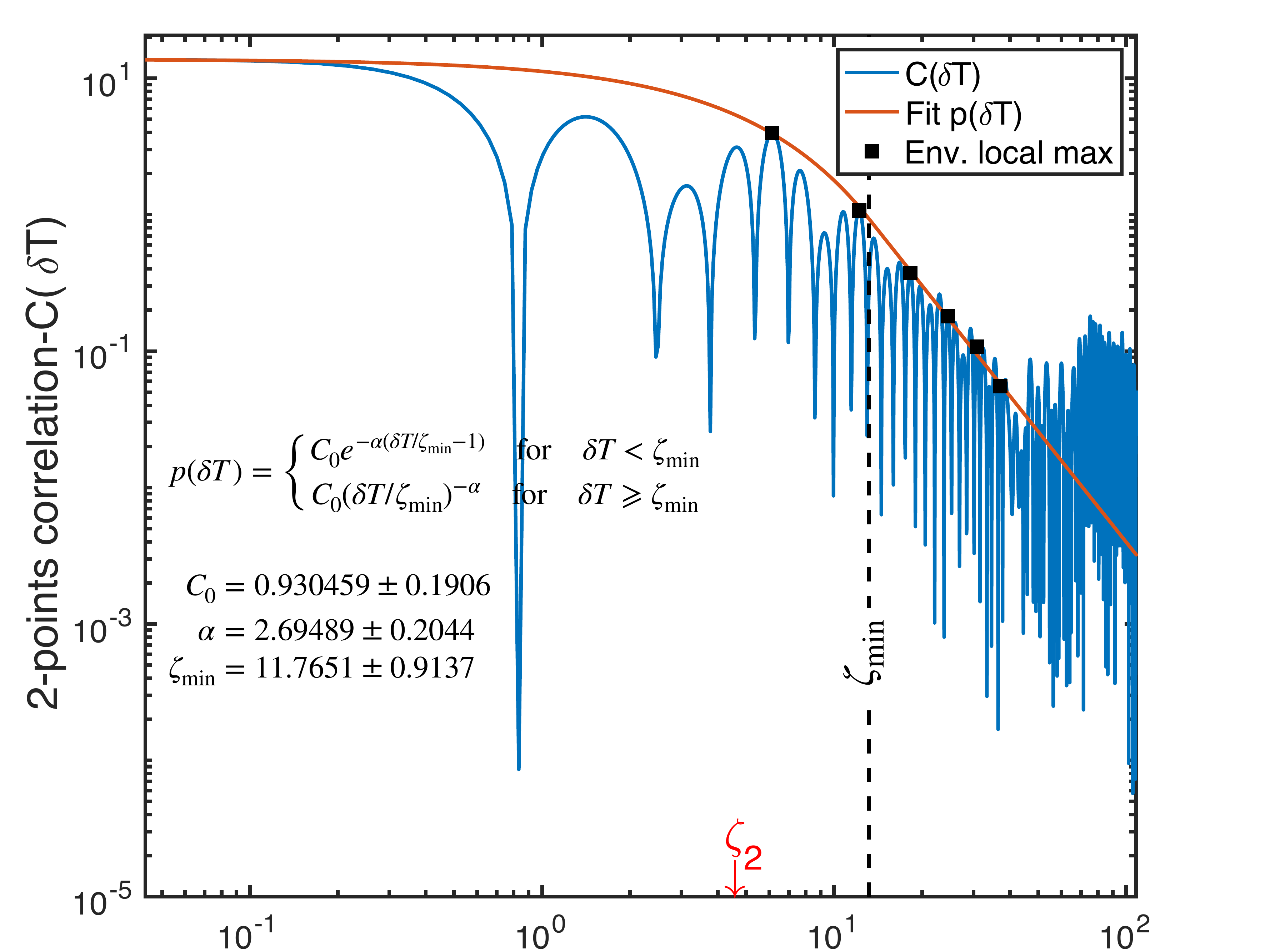}
	\caption{Equal time two-points correlation function in logarithmic scale. The red line correspond to the fit of the envelop with the function $p(\delta T) $  in inset. Black squares to the local maxima of the envelop.}
	\label{fig:loglogc2}
\end{figure}

\subsection{The mutual information vanishing range $\xi_I$}
Finite correlation range implies a vanishing range the information content shared by two locations separated by $\delta T$. Transfer entropy is set to determine the causality in the mutual information, it can be also use to estimate $\xi_I$. Setting the roundtrip lag at the location of P1$_1$ the profile of the transfer entropy is shown in Fig.~\ref{fig:loc_range}(c). It can also be seen that this quantity decays exponentially when the separation increases. We found that $\xi_I=4.3$ ps, which is of same order of $\xi_2$.
\begin{figure}
\centering
\includegraphics[width=.45\textwidth]{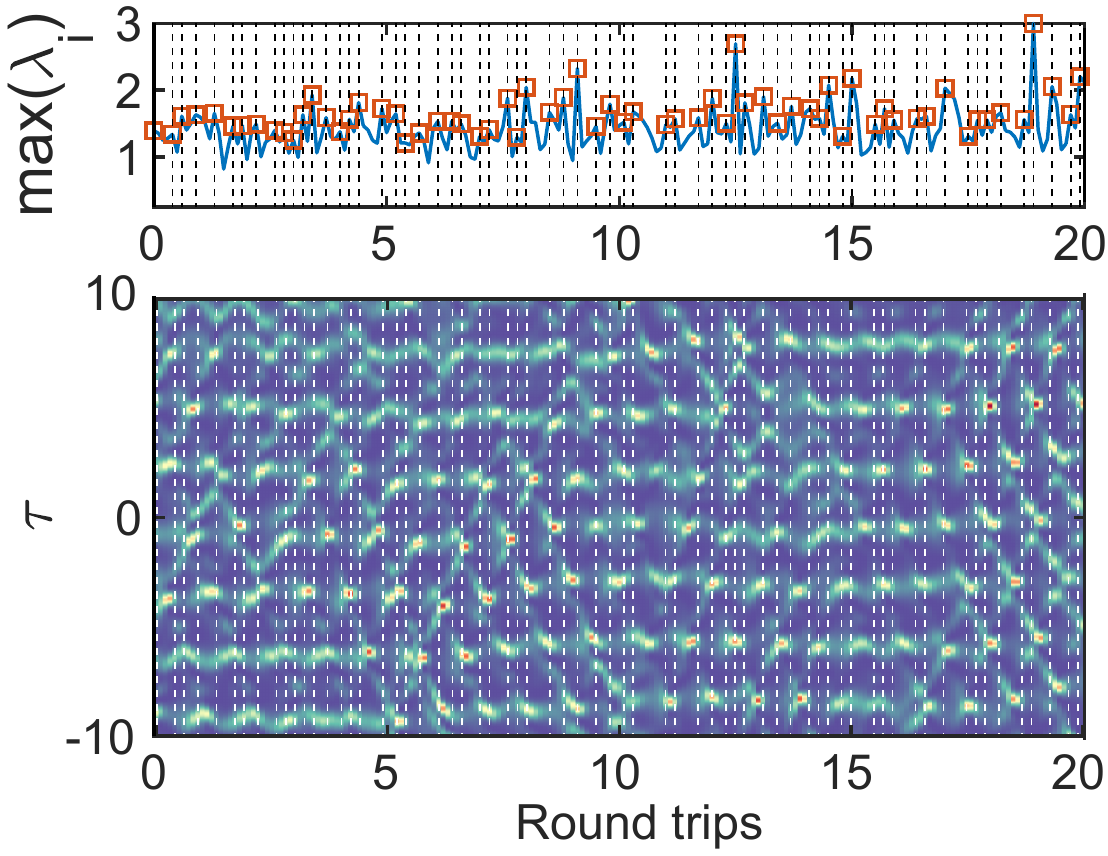}
\caption{Maximal local Lyapunov exponent (top panel) and the corresponding roundtrip evolution (bottom panel). Red square mark the local maxima of the largest Lyapunov exponents. Vertical lines give the local of the largest burst over the roundtrips.}
	\label{fig:loc_bursts_lyap}
\end{figure}
The three ranges clearly suggest that the correlations in the systems span beyond the chaotic subdomains. To understand the meaning of these quantities, we have represented the mean profile of the bursts in the unit of $\xi_\delta$ and $\xi_2=\xi_I$ in Fig.~\ref{fig:range_comp}. As it can be seen from this figure $\xi_\delta$ appears to be the maximal extension of the local chaotic objects that are the burst and  $\xi_2=\xi_I$ the range above which their information content vanishes. This sets $\xi_2=\xi_I$ as the best choice for the local dynamics range.

Notice that the assumption of exponential decay is based on the observation in many chaotic systems. However, in such a profile, it is also expected that data follows a power law starting from some as proposed in \cite{broido2019scale,clauset2009power}. We have analysed our data of $C\left(\delta T\right)$ according to this proposition. The result is given in Fig.~\ref{fig:loglogc2}. As excepted, this figure show a region where the envelop of $C\left(\delta T\right)$ follows a power, which is initially precede by a region where the distribution is exponential. The stating point of the power law is three times larger than the exponential decay range. In addition, the exponential decay range obtained following the reference \cite{clauset2009power}, is about 4.88, that is of  the same  order as the values that we have previously obtained by detecting the trend 2-points correlation function. Hence, in our analysis, the exponential regime allows us to determine the size of subdomains for the forecasting.
 
 \begin{table}
\centering
\begin{tabular}{c|c|cccc}
\toprule
\multicolumn{1}{c}{}&  \multicolumn{5}{c}{$\Delta T_p$ [ps]}\\
  \cmidrule(lr){3-6}
\multicolumn{1}{c}{} & \multicolumn{1}{c}{} & 22   & 45    & 180   & 360\\
     \cmidrule(lr){3-6}
\multirow{4}{*}{$(S/S_{th})^2$}  & 11 &  3.66  &  1.83 &   0.47  &  0.23\\
  &14  &  2.85  &  1.32  &  0.34 &   0.16\\
  &20 &  1.81   & 0.81  &0.21   & 0.10\\	
  &23 &  1.31    &0.71 &   0.18  &  0.09\\
\bottomrule
\end{tabular}
\caption{Lifetime of spatiotemporal chaotic fluctuations in fraction of the cavity roundtrip. $\Delta T_p$ is the duration of the pump temporal window and $(S/S_{th})^2$ is the ratio between the pump power $S^2$ and the threshold value $S_{th}^2$.}
\label{tab1:lifetine}
\end{table}

\subsection{Prediction horizon}
There are many definitions of the prediction time of chaotic evolution. These definition generally converge to the same value when dealing with low dimensional chaos with small number of positive Lyapunov exponents. The widely used is the Lyapunov time given $T_{\lambda_m}=1/\max(\lambda_i)$. However, this time has some limitations for high dimensional systems. In that case the maximal local Lyapunov exponent can be a good alternative. So far the bursts are the most chaotic objects in the system we can follow the local Lyapunov exponent together with the local evolution as shown in Fig.~\ref{fig:loc_bursts_lyap}. We can see that the local values of the maximum Lyapunov exponent correspond to the emergence of the at least one burst and  are much larger  than the mean value $\lambda_m = 0.8$ (see Fig.~\ref{fig2}(c). The Lyapunov time given by the mean value of these local feature is $T_{\lambda_l}\simeq 3$ roundtrips.
This time scale have to be compare with the Kolmogorov-Sinai entropy time \cite{dellago1997mixing,gualandi2020predictable}. This entropy is estimated as $h_{KS} = \sum\lambda_m$ with $\lambda_m$ the positive Lyapunov exponents. Then $T_{KS} = 1/h_{KS}$ give the time scale of entropy production by the system. We have computed this time for different pump temporal windows and pump intensity. The result is shown in Table~\ref{tab1:lifetine}. It appears that for our configuration this time is smaller than a roundtrip time.}
 
\section{Precursors-driven machine learning}
%\section{Results}

Since the spatiotemporal chaos generated in the resonator is a highly dimensional one, 
($\Delta T_{stc}\ll\Delta T_{p}$ and $\tau_{stc}<t_R$), forecasting the fully developed turbulence 
of the fiber ring cavity is a great challenge. Recent works using neural networks have opened new perspectives in this field~\cite{pathak2018model,amil2019machine,rafayelyan2020large,jiang2019model,vlachas2018data}. 
In particular, in~\cite{pathak2018model,vlachas2018data}, they have used an echo state network to reproduce 
the dynamics of the Kuramoto-Shivashinsky equation over several Lyapunov times. It  also shows that increasing 
the size of the system requires larger network nodes with the same forecasting accuracy that would be almost 
impossible in our system presenting a much higher spatiotemporal chaotic behavior compared to these works. 
Here, we propose to investigate an alternative to the forecast of the complete field under study. 
It consists of identifying a precursor of an event of interest and extracting a subdomain around it to reduce the complexity of the forecasting and increase the accuracy of the predictions.
For this purpose, we compute the information flow to optimize the determination of the size of {subdomains}. 
The transfer entropy 2D map is presented in Fig.~\ref{fig4:TE}(b) (numerics) and Fig.~\ref{fig4:TE}(g) {(experiments).  
At finite roundtrips, it exhibits either a central peak (P0$_i$) or double peak (P1$_i$) structures. As an example, the temporal profiles of  roundtrip 
lags P0$_1$ and P1$_1$, the most powerful, are shown in Fig.~\ref{fig4:TE}(a) and (f). Experimental traces are temporally wider because of the finite band-pass of the detection system with a noise background inherent to experiments, but a pretty good agreement with numerics is obtained}. These peaks {mean that, on average,} any peaks in the 
evolution {carry} information from its own past.  This information vanishes roundtrip to roundtrip 
(Fig.~\ref{fig4:TE}(b) and \ref{fig4:TE}(g)), the amplitude of the peaks following an exponential decay 
as can be seen in Fig.~\ref{fig4:TE}(c) and \ref{fig4:TE}(h). The dual peaks structure of the 
P1$_i$ has the advantage to be easily differentiated compared to the single peak of the P0$_i$ making P1$_i$ 
the better choice than bursts precursors. Furthermore, the time shift between the peaks of the P1$_i$ is of the 
same order of the equal time correlation range $\xi_2$ (see Appendix~\ref{ap:subdomains}) and can be appropriated 
to be the order of magnitude of our {subdomains}. Each measurement is locally centered at the location of the intensity burst. 
Given that information converges from P1$_j$ towards P0$_i$ we can make an association 
$\{$P0$_i$,P1$_j\}(|T-T_k|\leqslant \xi_2)$ and perform a supervised machine 
learning training, with $T_k$ being the location of the $k$-th local peak.

\begin{figure}[h]
\begin{centering}
\includegraphics[width=.4\textwidth]{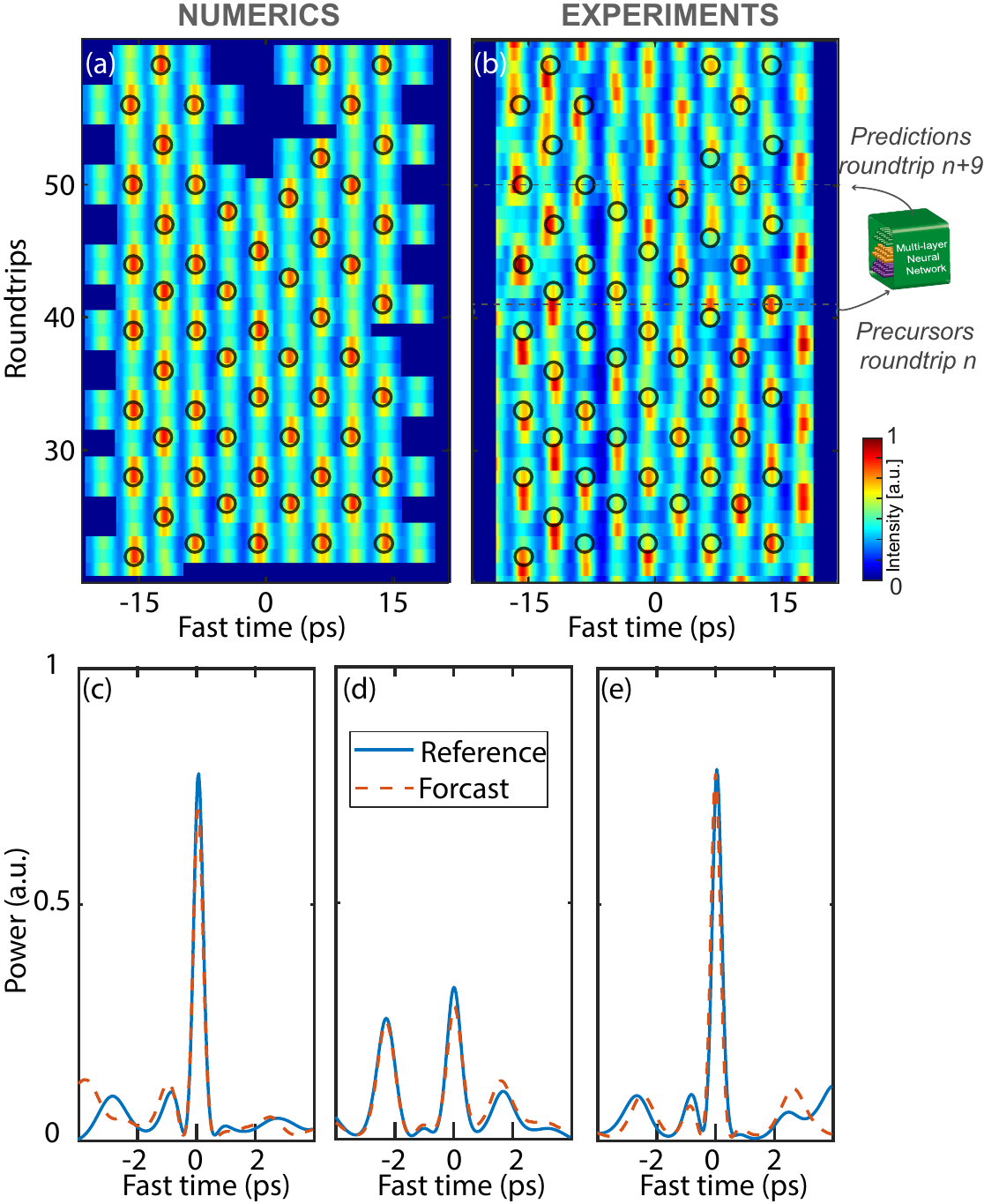}
\par\end{centering}
\caption{\label{fig5:accuracy} {Predictions of the \textit{when} and \textit{where} in circles, n+9 round-trips after a precursor is detected, superimposed on the spatiotemporal evolution of the field at the cavity output, (a) from numerics and (b) experiments. (c)-(e) Three typical examples of the \textit{what is coming?}: forecast pulse in orange dashed lines (predicted shape) to be compared to the reference one in blue solid lines. More comparison are provided as an additional material (plot\_pred\_anim.gif). The normalised parameters are $\Delta=1.1$ and $S=5.0$. Accuracy metrics give that 92$\%$ of the P1$_1$ precursors produce true positive prediction.}}
\end{figure}

%\section{Precursors-driven machine learning}
The network in Fig.~\ref{fig1:cartoon} is a deep Long Short-Term Memory (LSTM) encoder-decoder algorithm which {has} been shown to be suitable for sequence-to-sequence 
forecasting \cite{hochreiter1997lstm,lecun2015deep}. We used two sets of data: either two simulation runs or two experimental campaigns of recordings. One for training and testing, and the other one to evaluate fully independently the forecasting accuracy of our system.
Figures~\ref{fig4:TE}(d) and {\ref{fig4:TE}(e)} show the {highly} forecasting correlation skill on the test data for two roundtrip lags, P1$_1$, P1$_2$, from numerics, and figures~\ref{fig4:TE}(i) and {\ref{fig4:TE}(j)} from experiments. { The correlation is close to 97$\%$ for P1$_1$ and only decays to 60$\%$ for P1$_2$ in numerics proving the excellent performance of our method. In experiments (Figs.~\ref{fig4:TE}(i) and \ref{fig4:TE}(j)), while data are noisy, correlations factors remain still close to 1, with 75$\%$ and 64$\%$ for P1$_1$ and P1$_2$, respectively. }

{  After training and testing are completed,} we can now follow the evolution looking for the precursors. For the detection, we have used a moving window convolution of the current roundtrip with the profile of the P1$_i$. Once a double peak precursors   identified in the running roundtrip we feed of our multi-level network, and then we predict the positions of pulses that will appear \textit{m} {roundtrips} later. We performed the same process for numerical simulations and experiments. The {positions of reconstructed local predictions (\textit{when} and \textit{where})} are presented in circles in Figs.~\ref{fig5:accuracy}(a) and (b), superimposed on spatiotemporal traces of the output cavity. As illustrated on the {right-hand} side of Fig.~\ref{fig5:accuracy}(c), precursors are identified at roundtrip $n$ to forecast pulses which will appear at roundtrip  $n+9$. This is 3 times the horizon given by the local maximum Lyapunov exponent (see Supplemental information for the details).  We obtained a better accuracy in the predictions for numerical simulations (Fig.~\ref{fig5:accuracy}(b)) because experimental data (Fig.~\ref{fig5:accuracy}(c)) are { noisier}. The shape of the predicted pulses ("\textit{what is coming ?})" { is also predicted from our algorithm. Typical examples are depicted in Figs.~\ref{fig5:accuracy}(c) to \ref{fig5:accuracy}(e). An excellent agreement is achieved compared to the reference traces}. { The performances of the predictions in terms of false-positive and accuracy are summarized in Table 1. For P1$_3$, predictions at n+9 roundtrips is possible with less than $8\%$ of false-positive precursors while operating in a strongly chaotic spatiotemporal regime. At this horizon, the accuracy is about $60\%$. By slightly lowering the pump power from 3.3 to 3 times the cavity threshold (P1$_2$), the regime is still strongly chaotic. While we get the same ratio of false-positive prediction, we reach $75\%$ of 
prediction accuracy at n+6 roundtrips. For weakly spatiotemporal chaotic regimes, between two and three times the cavity threshold, the prediction are almost perfect.

\begin{table}
\begin{center}
 \begin{tabular}{ |c| c| c|c|}
   \hline
     & P1$_1$ & P1$_2$ & P1$_3$ \\
    \hline
   False-positive precursors  ($\%$) & 6.66 & 8 & 14.92 \\
    \hline
   True-positive precursors  ($\%$)& 74.60 & 71.30 & 66.4 \\
    \hline
  Pulses without precursors  ($\%$)& 25.40 & 28.70 & 33.6 \\
    \hline
   Prediction accuracy ($\%$) & 96.60 & 60 & - \\
    \hline
       Roundtrip forecast &3 & 9& 15 \\
    \hline
 \end{tabular}
\end{center}
\caption{Performances of the method as a function of the double peaks precursor (P1$_i$)}
\end{table}
}
 This value is remarkable for such a {high-dimensional} chaotic system. We also point out that all predictions based on P1$_1$  reach an accuracy above $90\%$ without any deep {optimization} of the network.

\section{Conclusion}
We have shown that forecasting high-dimensional chaos is possible by splitting the system around the object of interest. These objects are identified by mapping the transfer entropy. This quantity is also used to estimate the extension range 
of the subsystem, the precursors of the object of interest and also the forecasting horizon of the dynamics. By detecting the precursors, we can therefore build a triggerable model free process in which the "\textit{when}?" and "\textit{where}?" are no more the concerns but the "\textit{what is coming}?". Only the accuracy of this prediction is affected by the level of the transfer entropy.
{ Our analysis was based on the Long Short-Term Memory encoder-decoder algorithm. However, other methods can be implemented to recognize precursor-pulse pairs, such as gate recurrent unit, echo state network, and  deep learning. The optimal recognition method of precursor-pulse pairs is an open problem.}
Our protocol can be { applied} to any nonlinear {systems} independently from its size provided that information 
 flows are correctly computed.
%\bibliographystyle{naturemag}
%% Here is the endmatter stuff: Supplementary Info, etc.
%% Use \item's to separate, default label is "Acknowledgements"
\section*{Acknowledgements}
 SC, AM, FB acknowledge the LABEX CEMPI (ANR-11-LABX-0007) 
as well as the Ministry of Higher Education and Research, 
Hauts de France council and European Regional Development Fund (ERDF) 
through the Contract de Projets Etat-Region (CPER Photonics for Society P4S), 
the French government through the Programme Investissement d'avenir
(I-SITE ULNE/ANR-16-IDEX-0004 ULNE, projects EXAT, FUHNKC), Equipex T-REFIMEVE and 
H2020 Marie Sk\l odowska-Curie Actions (MSCA)(713694). M.G.C. thanks for the financial support of FONDECYT project {1210353} 
and Millennium Institute for Research in Optics, {ANID--Millennium Science Initiative Program-ICN17\_012.}
%\section{Competing Interests} The authors declare that they have no
%competing financial interests.
\section*{Author contributions}  A.M. and F.B. conceived, designed and performed the experiments. Numerical simulations based on the LLE were carried out by S.C. {and} M.G.C. The development of analytical tools and characterization of the ST chaos was performed by S.C. and M.G.C.. All authors participated in the analysis and interpretation of the results and the writing of the paper.
%\section*{Correspondence} Correspondence and requests for materials
%should be addressed to S.C\\
% (email: saliya.coulibaly@univ-lille.fr).

\appendix
%\subsection{}
%\subsection{}
\section{Experimental setup}
\label{ap:expsetup}
The experimental setup is based on a resonant passive fibre ring cavity mainly made of { a} 
132.9~$m$-long segment of highly nonlinear optical fibre (see supplementary material for details). 
The fibre has a low group-velocity dispersion of $\beta_2 = -3.8 $~ps$^2$km$^{-1}$ and a nonlinear 
coefficient $\gamma = 2.5$~W$^{-1}$km$^{-1}$ at $1545$ nm. The third-order dispersion ($\beta_3 \simeq 0.12$~ps$^3$km$^{-1}$) 
can be neglected (the {zero-dispersion} wavelength is located below 1500~nm far from the pump wavelength).  
A 95/5 input coupler is used to close the fibre loop cavity, whereas a 99/1 output coupler permits to extract and 
{analyze} the intracavity field. 
The output temporal field is {characterized} in real-time, round-trip to round-trip by means of a commercial 
time-lens system (Thorlabs, $300$ fs resolution over a $36$ ps time window). Details on the experimental 
setup and real time recording procedure are given in the supplemental document. 

\section{Numerical simulations}
\label{ap:numerics}
In the experimental setup, the propagation of light in {an} optical fiber loop is  {modeled} 
without loss of generality by the nonlinear Schr\"odinger equation augmented with boundary conditions or Ikeda map
 \cite{ikeda1979multiple,coen1997modulational,haelterman1992dissipative}:
\begin{eqnarray}
\partial_z A\left(z,T\right) &=&- i\frac{\beta_2}{2}\partial_T^2A\left(z,T\right)+i\gamma A\left(z,T\right)\left|A\left(z,T\right)\right|^2\label{eq:Ikedamap1}\qquad \nonumber \\
A\left(0,T+T_R\right)&=& \sqrt{\theta} E_i\left(T\right)+\sqrt{\rho} A\left(L,T\right)e^{-i\Phi_0}, \nonumber
\label{eq:Ikedamap2}
\end{eqnarray}
where $T_R$ stands for the round-trip time, which is the time taken by the pulse to propagate along the 
cavity with the group velocity, $\Phi_0$ is the linear phase shift, $\theta\ \left(\rho\right)$ is the mirror 
transmission (reflection) coefficient, and $L$ is the cavity length. The complex amplitude of the electric 
field inside the cavity is $A$. Each of the coefficients $\beta_2$ is responsible for the {second-order }
dispersion,  and $\gamma$ is the nonlinear coefficient of the fiber. The independent variable $z$ refers to 
the longitudinal {coordinate,} while $T$ is the time in a reference frame moving with the group velocity 
of the light. For large enough cavity finesse $\mathcal{F=\pi/\alpha}$, with $\alpha$ the effective 
losses of the cavity, the evolution of the electric field inside the loop is well described by the 
Lugiato-Lefever equation  \cite{lugiato1987spatial,haelterman1992dissipative}:
\begin{equation}
\frac{\partial \psi}{\partial t}=S-(1+i\Delta)\psi-i\eta\frac{\partial^{2} \psi}{\partial \tau^{2}}+i\vert\psi\vert^{2}\psi,
\label{eq:LL}
\end{equation}
{where} $S=2E_i\sqrt{\gamma L/\alpha^3}$, $\psi=A\sqrt{\gamma L/\alpha}$, 
$t=\alpha T/T_R=\alpha m$, 
and $\tau=T/T_n$ { with $T_n=\sqrt{\left|\beta_2L\right|/(2\alpha)}$}. $\delta=(2k\pi-\Phi_0)/\alpha$ is the detuning 
with respect to the nearest cavity resonance $k$.  The integer $m$ gives the roundtrip number and the coefficient $\eta=\pm 1$ is the sign of the 
group velocity dispersion term. The configuration of our setup gives $\alpha=0.20$, $\Delta=1.1$ and $T_n\simeq 1.1$~ps.

\section{Lyapunov spectrum computation}
\label{ap:lyapunov}
Strictly speaking, to prove a spatiotemporal chaotic dynamics, one may  compute several quantities. 
In particular, it is mandatory to compute the Lyapunov spectrum. Next, this spectrum must have 
a positive part and continuous region whose area has to linearly increase with the 
size of the system. The computation of the Lyapunov spectrum itself is very well 
documented \cite{skokos2010lyapunov} and is not the purpose here. Let just recall the main steps.
From the state of the system at a given time, the linear evolution of any small perturbation $\delta\textbf{X}$
can be described by  $\partial_t\delta\textbf{X}=\textbf{J}\delta\textbf{X}$, where  $\textbf{J}$ { is} 
the respective Jacobian. {In the present case, we introduce $\psi=\psi_r+i \psi_i$, with $\psi_r$ and $\psi_i$ being the real and imaginary part of $\psi$ respectively. At a time $t=t_0$, introducing $\psi=\psi_0+\delta\psi$, with $\delta\psi\ll\psi(t=t_0)=\psi_0$ the matrix \textbf{J} reads : 
\begin{equation}
\textbf{J} = \begin{bmatrix}
-(\alpha+2{\psi_0}_r {\psi_0}_i)&\quad&\delta - {\psi_0}^2_r - 3{\psi_0}^2_i-\partial^2_\tau\\
\\
-\delta + {\psi_0}^2_i+ 3{\psi_0}^2_r)+\partial^2_\tau&\quad&-(\alpha-2{\psi_0}_r {\psi_0}_i)
\end{bmatrix},
\end{equation}
and $\delta\textbf{X}=(\delta\psi_r,\delta\psi_r)^t$.
}
Suppose that we want to compute the $n$-th first { dominant exponents} 
of the spectrum, we introduce the matrix $\textbf{L}$, that contains $n$ orthonormal vectors $\textbf{v}_i$ which { to be used as initial conditions when solving $\partial_t\delta\textbf{X}=\textbf{J}\delta\textbf{X}$:}
\begin{equation}
\textbf{L}\left(t= t_0\right)\equiv\left[\textbf{v}_1\quad \textbf{v}_2\quad\dots \quad\textbf{v}_n\right]=\begin{bmatrix}
    x_{11}       & x_{12} & x_{13} & \dots & x_{1n} \\
    x_{21}       & x_{22} & x_{23} & \dots & x_{2n} \\
    \hdotsfor{5} \\
    x_{d1}       & x_{d2} & x_{d3} & \dots & x_{dn}
\end{bmatrix},
\end{equation}
{where $d$ is the dimension of the system.  After a time increment $dt$, the matrix 
$\textbf{L}$ evolves to $\textbf{L}\left(t_0+dt\right) = \hat{\textbf{U}}\textbf{L}\left(t_0\right)$ 
where $\hat{\textbf{U}} = e^{\textbf{J}*dt}$. Using the modified Gram-Schmidt \textbf{QR} 
decomposition on $\textbf{L}\left(t_0+dt\right)$, the diagonal elements of $\textbf{R}$ 
account for the Lyapunov exponents $\tilde{\lambda}_i\left(i=1,\dots, n\right)$ at time $t_0+dt$, that is}
\begin{eqnarray}
\tilde{\lambda}_i(t_0+dt) = \frac{1}{dt}\ln\left(\textbf{R}_{ii}(t_0+dt)\right).
\end{eqnarray}
Repeating this procedure several time, after a large number of iterations $N$, the Lyapunov exponents can be approximated by
\begin{eqnarray}
\lambda_{i}\equiv\langle\tilde{\lambda}_i\rangle = \frac{1}{Ndt}\sum_{k = 1}^{N}\ln\left(\textbf{R}_{ii}(t_0+kdt)\right).
\end{eqnarray}
From the spectrum $\left\{\lambda_i\right\}$ an estimator of the dimension of the chaotic attractor is given by
the {Kaplan-Yorke} dimension $D_{KY}=p+\sum_{i=1}^p\lambda_i/|\lambda_{p+1}|$ 
where $p$ is such that $\sum_{i=1}^p\lambda_i>0$ and $\sum_{i=1}^{p+1}\lambda_i<0$ \cite{ott2002chaos}. 
For a {one-dimensional} system of size $L$, a spatiotemporal chaos implies that $D_{KY}$ increase linearly with $L$. 

\section{Determination of the subdomains to forecast}%Spatiotemporal chaos dimension  $\xi_\delta$
\label{ap:subdomains}
 In a spatiotemporal chaotic system many quantities can be used as order parameter. Considering the extensive feature of this chaos, the {Kaplan-Yorke} dimension 
may change linearly with the volume of the system \cite{Ruelle1982,cross1993pattern}. {Namely}, for a 1D system, $D_{YK}=\xi_\delta^{-1} \Delta T$ 
where $\Delta T$ {is} the extension of the system and $\xi_\delta$ represents the dimension correlation length of the system 
for a {fixed} value of the control parameter. This is an intensive quantity which gives an estimation of the extension of the dynamically independent subsystems.
Together with the dimension correlation length one can compute the correlation length $\xi_2$.
%\subsection{Equal time correlation dimension $\xi_2$}\ \\
This length is defined as the exponential decay range of the equal time two-point correlation \cite{Ohern1996lyapunov,egolf1994relation,cross1993pattern}:
\begin{equation}
\label{eq:2ptcorr}
C\left(\Delta\tau\right)=\langle\left(\psi\left(\Delta\tau+\tau^\prime,t\right)-\langle\psi\rangle\right)\left(\psi\left(\tau^\prime,t\right)-\langle\psi\rangle\right)\rangle,
\end{equation}
where the brackets  $\langle\cdot\rangle$ stand for the average process. The direct determination of $C\left(\Delta\tau\right)$ is quite costly in calculation time. However, by using the Wiener-Khintchin theorem \cite{reif2009fundamentals,egolf1995characterization}, it is computed by the following process: first time-averaging the Fourier spectra and next taking the inverse Fourier transform {of} its magnitude squared. Since the experimental spectra result from an averaging process over a large number of cavity roundtrip, $C\left(\Delta\tau\right)$ can also be computed taking the inverse Fourier transform of the measured spectrum. 
Hence, for the LLE {(\ref{eq:LL})}, we have computed $\xi_\delta$ end $\xi_2$ with respect to the input pump intensity. 
The third length we have computed is the long range decay rate of the transfer entropy $\xi_{TE}$. The region around the burst to forecast is largest range between $\xi_\delta$, $\xi_2$ and $\xi_{TE}$. Detailed implementation can be found in the SI.

\section{The forecasting protocol}
\label{ap:predprotocol}
%\begin{figure}
\begin{algorithm}[H]
 \caption{protocol}
 \begin{algorithmic}
 \State  \bf{From data:} \normalfont 
    \begin{itemize}
        \item Compute the transfer entropy map
        \item Detect the pulses
        \item Move backward to the chosen history
     \end{itemize}
 \State  \bf{The training:} \normalfont 
            \begin{itemize}
                \item compute the PDF from peaks amplitude
                \item split data 80\% for training and 20\% for testing with the same PDFs
                \item Standardize input data (precursors) : Yeo-Johnson transform
                \item Create the LSTM encoder-decoder (see Table~\ref{tab:model-summary})
            \end{itemize}
 \State  \bf{Forcasting:} \normalfont 
            \begin{itemize}
                \item Watch the dynamics for precursors
                \item If precursor detected feed the network to forecast the incoming pulse at the chosen horizon and location 
                \end{itemize}
\end{algorithmic}
\end{algorithm}
%\end{figure}

\begin{table} 
\begin{tabular}{lll} 
Model: \\ \hline 
Layer (type)                   & Output Shape                & Param \#    \\ \hline \hline 
lstm\_1 (LSTM)                 & (None, 3, 820)              & 3365280    \\ \hline 
lstm\_2 (LSTM)                 & (None, 820)                 & 5382480    \\ \hline 
repeat\_vector (RepeatVector) & (None, 1, 820)              & 0          \\ \hline 
lstm\_3 (LSTM)                 & (None, 1, 820)              & 5382480    \\ \hline 
lstm\_4 (LSTM)                 & (None, 1, 820)              & 5382480    \\ \hline 
time\_distributed\_10 (TimeDistr) & (None, 1, 205)              & 168305     \\ \hline \hline 
Total params: 19,681,025 \\ 
Trainable params: 19,681,025 \\ 
Non-trainable params: 0 \\ \hline 
\end{tabular} 
\caption{Python LSTM encoder-decoder summary for the network we have trained with linear activation function. We have used the package Tensorflow-Keras. For the optimization with have used the following options: "optimizer=Adam(), loss='Huber',metrics='mae'".  } 
\label{tab:model-summary} 
\end{table}
%
% TABLES
%
% If there are any tables, put them here.
%
\bibliographystyle{unsrt}

%\bibliography{\jobname}

\end{document}